\documentclass[%
 reprint,
 amsmath,amssymb,
 aps,prd
]{revtex4-1}

\usepackage{graphicx, color}
\usepackage{dcolumn}
\usepackage{bm}

\newcommand{\Fermi}{\textit{Fermi}}
\newcommand{\tuc}{47 Tuc}

\begin{document}


\title{Comment on "Understanding the $\gamma$-ray emission from the globular cluster 47 Tuc: evidence for dark matter?"}

\author{Richard Bartels}
 \email{r.t.bartels@uva.nl}
\author{Thomas Edwards}%
 \email{t.d.p.edwards@uva.nl}
\affiliation{%
Gravitation Astroparticle Physics Amsterdam (GRAPPA),\\
Institute for Theoretical Physics Amsterdam
and Delta Institute for Theoretical Physics,\\
University of Amsterdam, Science Park 904, 1098 XH Amsterdam, The Netherlands
}%




\date{\today}

\begin{abstract}
In a recent paper Ref.~\cite{Brown:2018pwq} analyzed
the spectral properties of the globular cluster 47 Tucanae (47 Tuc) using 9 years of \Fermi-LAT data.
Ref.~\cite{Brown:2018pwq} argues that the emission from 47 Tuc cannot be explained by millisecond pulsars
(MSPs)
alone because of a significant discrepancy between the MSP spectral properties and those
of 47 Tuc. It is argued that there is a significant ($>5\sigma$) preference for a two source scenario.
The second component could be from the annihilation of dark matter in a density spike surrounding 
the intermediate-mass black hole candidate in 47 Tuc.
In this paper we argue that the claimed discrepancy arises because Ref.~\cite{Brown:2018pwq} uses a stacked MSP spectrum to model the emission from MSPs in \tuc which is insufficient to account
for the uncertainties in the spectrum of the MSPs in \tuc.
Contrary to the claims in Ref.~\cite{Brown:2018pwq}, we show that the significance of an additional
dark matter component is $\lesssim 2\sigma$ when sample variance in the spectrum of a population 
of MSPs is accounted for. The spectrum of \tuc\ is compatible with 
that of a population of MSPs similar to the disk population. 
\end{abstract}

\maketitle


\section{Introduction}
\label{sec:intro}
Globular clusters (GCs) are old ($\sim 10 \mathrm{\,Gyr}$) systems typically hosting
a sizable population of millisecond pulsars (MSPs). $\gamma$-ray emission from 
GCs is expected to be dominated by their MSP population \cite{2010A&A...524A..75A}.
For certain GCs the $\gamma$-ray emission can receive a large contribution 
from only one or a few pulsars
\cite{2011Sci...334.1107F, Johnson:2013uza,Hooper:2016rap}. On the other hand, for \tuc\ the $\gamma$-ray
emission appears to arise from a larger number of MSPs \cite{Abdo:2009rjc}.

In a recent paper, Ref.~\cite{Brown:2018pwq} analyzed the $\gamma$-ray emission from
\tuc\ using 9 years of \Fermi-LAT data. The obtained spectrum of \tuc\
is then compared to the stacked MSP spectrum from Ref.~\cite{Xing:2016ncp}. 
It is argued that the stacked MSP spectrum provides a poor fit to
the data and a second emission component is required. Ref.~\cite{Brown:2018pwq} then
includes a dark matter (DM) signal, $\chi\chi\rightarrow b\bar{b}$, 
which could arise from annihilation of dark matter in a density spike around the 
candidate intermediate-mass black hole present in \tuc \cite{Brown:2018pwq}.
The combined DM + MSP model significantly improves the fit to the spectrum of \tuc\ by
a $\Delta TS = 40$ compared to stacked MSPs only.
This discrepancy between the observed spectrum of \tuc\ and that of stacked MSPs
is then interpreted as possible evidence for DM within a globular cluster \cite{Brown:2018pwq}.
It is pointed out by Ref.~\cite{Brown:2018pwq} that since the stacked spectrum
of MSPs is determined from observations of local MSPs, the conclusions could be altered
if the spectral properties of MSPs in \tuc\ differ from those of the local population.
However, based on the comparison of disk and GC MSPs at other wavelengths the authors
deem this unlikely \cite{Brown:2018pwq, Abdo:2009rjc, Abdo:2009jfa, Grindlay:2001xh}.

In this paper, we show
that the spectrum of \tuc\ is well explained by a population of MSPs
with similar properties to the MSPs in the disk, contrary 
to the conclusions of Ref.~\cite{Brown:2018pwq}. 
Our argument relies on the fact that the stacked MSPs 
spectrum used by Ref.~\cite{Brown:2018pwq} does not
adequately represent the uncertainty in the spectrum 
of the MSP population in \tuc. When these uncertainties
are accounted for the significance of the need for a second emission component drops drastically .
Consequently, there is
no discrepancy between the spectrum of \tuc\ and that of a population of MSPs, and 
thus no potential evidence for dark matter.

\section{\label{sec:methods}Methods}
We wish to assess the probability of obtaining data as extreme of more extreme as that observed when considering the MSP only hypothesis.
In this way we can assess whether there is a significant hint of a DM signal in 
the $\gamma$-ray spectrum of \tuc.
For this purpose, we
simulate $10^{4}$ mock $\gamma$-ray spectra representative of \tuc\ assuming all emission arises from MSPs.
For consistency with Ref.~\cite{Brown:2018pwq} we assume that the characteristics of the
MSPs in \tuc\ are similar to those in the Galactic disk. 
First, we sample MSPs until we saturate the luminosity of \tuc\ 
($L_\mathrm{\tuc} = (6.45\pm0.19)\times 10^{34} \mathrm{\,erg\,s^{-1}}$)
from $0.1$--$100\mathrm{\,GeV}$ (note that the luminosity of \tuc\ is comparable to
that of the brightest MSPs).
Luminosities are drawn randomly from the best-fit lognormal luminosity function for
disk MSPs
obtained by Ref.~\cite{Bartels:2018xom}, 
\mbox{$dN/dL \propto L^{-1}\exp\left[-\left(\log_{10} L - \log_{10} L_0\right)^2/2\sigma_L^2\right]$},
with $\sigma_L = 0.63$ and $L_0 = 4.1\times10^{32} \mathrm{\,erg\,s^{-1}}$\footnote{Similar results were found using a broken power law.}.
We then assign each MSP a spectrum, which is sampled with replacement from the spectra
of the 39 MSPs in the second \Fermi-LAT catalog of $\gamma$-ray pulsars (2PC) \cite{TheFermi-LAT:2013ssa}. For maximum consistency we use the spectra as derived by Ref.~\cite{Xing:2016ncp}.
We note that there is no obvious correlation between the $\gamma$-ray spectra 
and the MSP luminosity, justifying our choice to sample spectra and luminosities independently.
However, for completeness we mention that assigning to each spectrum the corresponding luminosity from the 2PC (determined
using distance proxies as in Ref.~\cite{Bartels:2018xom}), rather than a random luminosity, 
does not alter the results discussed below.

After having created $10^4$ representative $\gamma$-ray spectra of the MSP population in 
\tuc\ we fit these mock spectra with 
(i) the stacked MSP spectrum from Ref.~\cite{Xing:2016ncp},
and 
(ii) including an
additional dark matter spectrum corresponding to
a $34\mathrm{\,GeV}$ particle
annihilating into $b\bar{b}$ \cite{Cirelli:2010xx}. 
This analysis is similar to what has been performed by \cite{Brown:2018pwq} on
the true spectrum of \tuc.
Our mock data, representing random MSP populations that could be present in \tuc,
allows us to study how likely it is that we find a preference for a DM component
when fitting this population with the stacked MSP spectrum from
Ref.~\cite{Xing:2016ncp}.
In order to test by how much an additional DM component improves the fit over a stacked
MSP spectrum only, we 
perform a Gaussian likelihood analysis minimizing 
the $\chi^2$ test statistic:
$\chi^2 = \sum_i^N (\mu_i - \left.E^2dN/dE\right|_\mathrm{i,\,MC})^2 / \sigma_i^2$.
Here the data points $\left(\left.dN/dE\right|_\mathrm{i,\,MC}\right)$ come
from our Monte-Carlo realizations of MSP populations and the errors ($\sigma_i$) 
are chosen such that the fractional errors are identical to those in Fig.~1 from Ref.~\cite{Brown:2018pwq}
(also see Fig.~\ref{fig:spectrum} of this work). 
The expectation ($\mu$) 
refers to (i) the stacked MSP spectrum, or (ii) the stacked MSP + DM spectrum for a 
$34\mathrm{\,GeV}$ particle annihilating into $b\bar{b}$. We sum
over all bins with the exception of those containing an upper-limit, i.e.~$N = 11$.
Models are compared by computing $\Delta \mathrm{TS}$.
Applying this analysis to the actual data from \tuc\, instead of simulated data, yields $\Delta \mathrm{TS}=36$. We thus agree with Ref.~\cite{Brown:2018pwq}
that there appears to be a significant discrepancy when
the stacked MSP spectrum is used to fit the emission from \tuc\ (see Fig.~\ref{fig:spectrum}). 

We note that the $\Delta \mathrm{TS}=36$ obtained in this work is slightly smaller than that obtained by
Ref.~\cite{Brown:2018pwq} ($\Delta \mathrm{TS}=40$). Most likely this is due to a number of differences in the analyses. Ref.~\cite{Brown:2018pwq} uses 10 logarithmic bins per decade, instead of the 5 logarithmic bins per decade used in this work. Moreover, Ref.~\cite{Brown:2018pwq} fit the MSP and MSP + DM spectra to \tuc\ using
the \texttt{Fermi Science Tools} \cite{2017arXiv170709551W}, whereas we perform a Gaussian likelihood analysis using the (simulated) spectrum of \tuc\ directly.
However, since the difference in $\Delta \mathrm{TS}$ obtained for the spectrum of \tuc\ between our work and that of Ref.~\cite{Brown:2018pwq} is small, we do not expect any qualitative effect
on the conclusion of this work due to differences in the analyses.

\section{\label{sec:results}Results}
\begin{figure}[h]
\includegraphics[width=\linewidth]{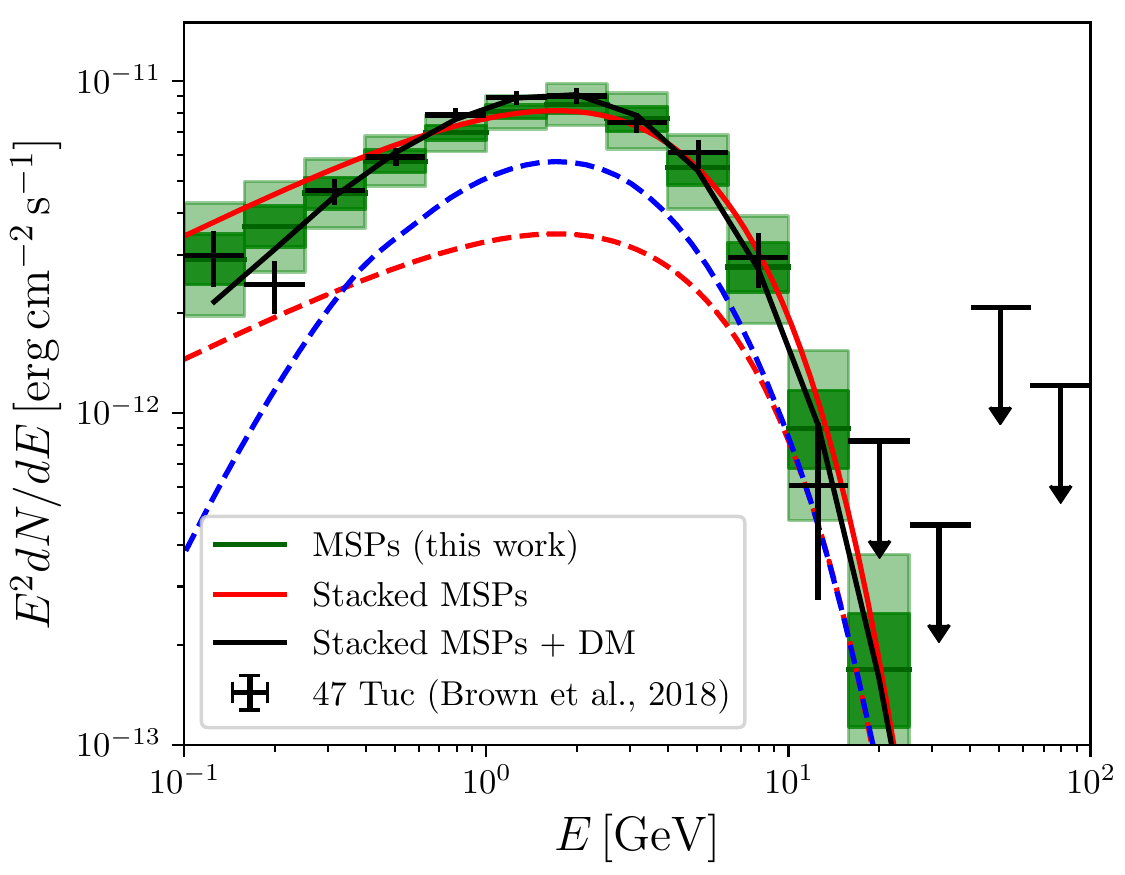}
 \caption{\label{fig:spectrum}Spectrum derived for $10^4$ mock realizations of 47 Tuc is
 shown in green. The dark (light) green bands show the 68\% (95\%) containment interval.
 The spectrum of \tuc\ as derived by Ref.~\cite{Brown:2018pwq} is shown by the black data points.
 The solid-red line shows the best-fit stacked MSP spectrum and the solid-black line
 the best-fit stacked MSP (dashed red) + DM (dashed blue) spectrum.}
\end{figure}

On average our realizations of \tuc\ contain $\mathcal{O}(50)$ MSPs, but this can
differ by a factor $\sim 2$ for individual realizations. These numbers are compatible with
the observed number of MSPs in \tuc\ \cite{Freire:2017mgu}. Half of the flux is
typically contributed by the 5 brightest sources.
In Fig.~\ref{fig:spectrum} we show the spectrum of \tuc\ as derived by 
Ref.~\cite{Brown:2018pwq} (black errorbars) and the MSP spectra derived in
our simulations (green). The dark green line is the median, with the dark (light)
green band representing the 68\% (95\%) containment interval.
Within errors, the simulated MSP spectra agrees with the observed spectrum in 
\tuc.
For reference we also show the best-fit stacked MSP spectrum (solid red).
The black solid line shows the best-fit stacked MSP (dashed red) + DM spectrum 
(dashed blue) spectrum.

\begin{figure}[t]
  \includegraphics[width=\linewidth]{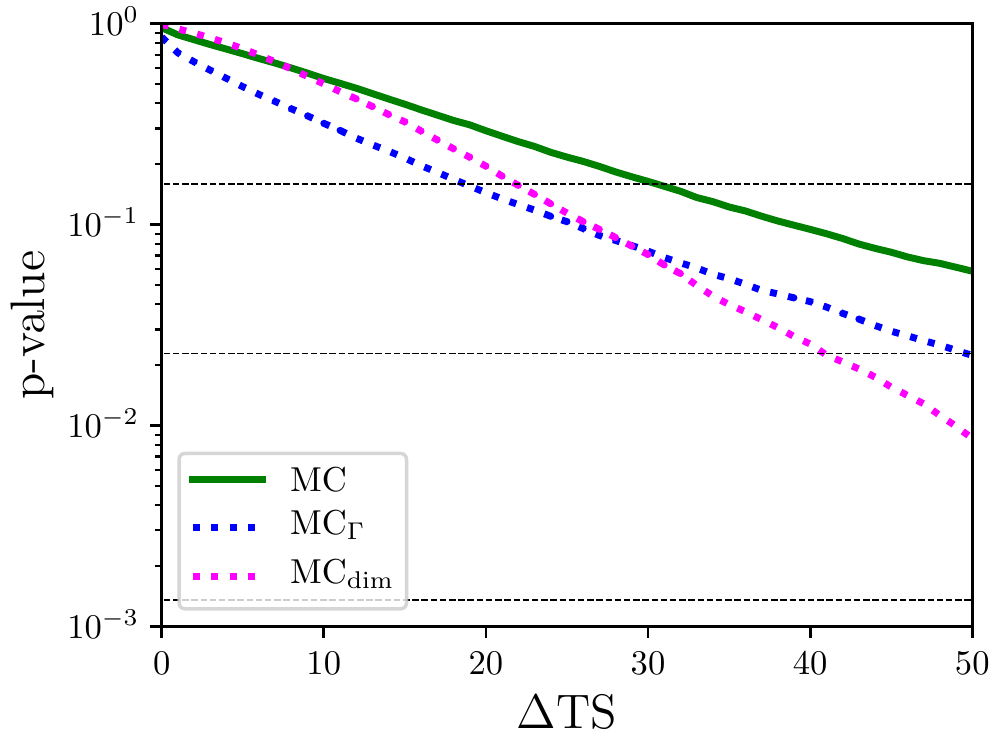}
   \caption{\label{fig:TS}
   p-value for a particular 
   $\Delta \mathrm{TS}$ when a DM signal, corresponding to a $34\mathrm{\,GeV}$
   particle annihilating to $b\bar{b}$, is included on top of the stacked 
   MSP spectrum of Ref.~\cite{Xing:2016ncp}.
   The green line is for our benchmark simulation in which we sample luminosities from
   a log-normal luminosity function \cite{Bartels:2018xom} and spectra are drawn from
   the 39 sources in the 2PC \cite{Xing:2016ncp}. The dotted-blue line is derived
   only sampling spectra from the brightest sources. Then dotted-magenta line
   corresponds to a scenario in which no single MSP contributes more than 10\% to the
   total flux of \tuc. Finally, the black-dashed horizontal lines correspond 
   to $1, 2, 3\sigma$.
   }
\end{figure}

Figure \ref{fig:TS} shows the p-value for a particular $\Delta \mathrm{TS}$ improvement in the fit to the mock spectra
of \tuc\ when a dark matter component, characterized by $M_\mathrm{DM}=34\mathrm{\,GeV}$ and annihilation into $b\bar{b}$, is included 
on top of the stacked MSP spectrum from Ref.~\cite{Xing:2016ncp}.
The results can be interpreted in a frequentist fashion. 
Let the null hypothesis ($H_0$) be MSPs only. The alternative hypothesis ($H_1$)
represents a need for dark matter with fixed mass. We choose between the two models by fitting a stacked MSP spectrum and the same spectrum plus a DM signal to the data.
Ref.~\cite{Brown:2018pwq} reports $\Delta \mathrm{TS}=40$ for the data from \tuc. From Fig.~\ref{fig:TS} we find that this corresponds
to a p-value of $p=0.09$ (solid green line), or $1.3\sigma$
\footnote{The conversion from p-value ($\alpha$) to $\sigma$ is performed
using the inverse-cumulative-distribution function for a standard-normal distribution:
$\mathrm{CDF}^{-1}_\mathcal{N}\left(1-\alpha\right)$.}.
In other words, there is no significant evidence for an additional, DM-like, component.

We repeated the above exercise only sampling spectra from the 16 brightest sources, which have a flux $F\geq 2\times 10^{-11}\mathrm{\,erg\,cm^{-2}\,s^{-1}}$ above $100 \mathrm{\,MeV}$. For the brightest sources we expect the most accurate determination of the spectrum.
The average 
spectra of these sources is slightly softer than for the full sample.
This analysis yielded the blue-dotted line in Fig.~\ref{fig:TS}.
In addition, we also tested for scenarios in which the MSPs 
in \tuc\ are unusually dim by requiring no single MSP to contribute more than 
10\% to the total flux of \tuc\ (dotted magenta).
In these analyses the findings of Ref.~\cite{Brown:2018pwq} correspond
to $1.8\sigma$ and $2.0\sigma$, respectively. Our conclusions thus remain
the same.

Our analysis shows that a stacked MSP spectrum, as applied by Ref.~\cite{Brown:2018pwq}, is insufficient to account for the variation in spectra that can come from different MSP populations. 
We have shown that in case of the data of \tuc\ an additional DM component is
not required.
Finally, we note that we fixed the dark matter mass to $34\mathrm{\,GeV}$.
Letting the mass
float can only shift the lines in Fig.~\ref{fig:TS} further to the right and
thus would imply even larger p-values for $\Delta \mathrm{TS}$, strengthening our conclusions. 

\section{Conclusion}
We simulated mock data for a population of MSPs, with characteristics similar
those in the Milky-Way disk, representative of that in \tuc. It
was shown that if the spectra of these populations are fit with the stacked MSP spectrum from Ref.~\cite{Xing:2016ncp} an additional DM spectrum corresponding to 
$M_\mathrm{DM}=34\mathrm{\,GeV}$ and annihilation into $b\bar{b}$ typically improves
the fit, because the stacked MSP spectrum does not capture the variation in the spectra of different MSP populations. 
Using our simulations we were able to assign a p-value to findings of 
Ref.~\cite{Brown:2018pwq}, who found $\Delta \mathrm{TS}=40$ improvement when including
DM, instead of a stacked MSP spectrum only, in a fit to the actual spectrum of \tuc.
We find that under all circumstances $\Delta \mathrm{TS}=40$ has a high-probability of
occurring and does not imply potential evidence for an additional spectral component ($\lesssim 2 \sigma$).
Rather, the discrepancy claimed by Ref.~\cite{Brown:2018pwq} arises because they represent the MSP-induced $\gamma$-rays in \tuc\ by
a stacked spectrum of MSPs. 
Stacked spectra are constructed to 
provide a proper fit to all MSP spectra simultaneously \cite{Xing:2016ncp}, 
but do not necessarily reflect accurately the total emission from a
population of MSPs, which can show large variations even when coming from 
a few dozen of sources.
We conclude that
there is no evidence for dark matter from the globular cluster \tuc.

\section*{Acknowledgments}

We thank the corresponding author of Ref.~\cite{Brown:2018pwq} for a responsive dialogue. We thank Daniele Gaggero, Jason Hessels, Dan Hooper, Bradley Kavanagh, 
Tim Linden, Chris McCabe, Emma Storm and Christoph Weniger 
for discussion.
This research is funded by NWO through 
a GRAPPA-PhD fellowship (022.004.017; RB)
and through the VIDI research program
"Probing the Genesis of Dark Matter" (680-47-532 - PI: Christoph Weniger; TE).


\bibliographystyle{apsrev4-1}
\bibliography{lib}

\begin{thebibliography}{16}%
\makeatletter
\providecommand \@ifxundefined [1]{%
 \@ifx{#1\undefined}
}%
\providecommand \@ifnum [1]{%
 \ifnum #1\expandafter \@firstoftwo
 \else \expandafter \@secondoftwo
 \fi
}%
\providecommand \@ifx [1]{%
 \ifx #1\expandafter \@firstoftwo
 \else \expandafter \@secondoftwo
 \fi
}%
\providecommand \natexlab [1]{#1}%
\providecommand \enquote  [1]{``#1''}%
\providecommand \bibnamefont  [1]{#1}%
\providecommand \bibfnamefont [1]{#1}%
\providecommand \citenamefont [1]{#1}%
\providecommand \href@noop [0]{\@secondoftwo}%
\providecommand \href [0]{\begingroup \@sanitize@url \@href}%
\providecommand \@href[1]{\@@startlink{#1}\@@href}%
\providecommand \@@href[1]{\endgroup#1\@@endlink}%
\providecommand \@sanitize@url [0]{\catcode `\\12\catcode `\$12\catcode
  `\&12\catcode `\#12\catcode `\^12\catcode `\_12\catcode `\%12\relax}%
\providecommand \@@startlink[1]{}%
\providecommand \@@endlink[0]{}%
\providecommand \url  [0]{\begingroup\@sanitize@url \@url }%
\providecommand \@url [1]{\endgroup\@href {#1}{\urlprefix }}%
\providecommand \urlprefix  [0]{URL }%
\providecommand \Eprint [0]{\href }%
\providecommand \doibase [0]{http://dx.doi.org/}%
\providecommand \selectlanguage [0]{\@gobble}%
\providecommand \bibinfo  [0]{\@secondoftwo}%
\providecommand \bibfield  [0]{\@secondoftwo}%
\providecommand \translation [1]{[#1]}%
\providecommand \BibitemOpen [0]{}%
\providecommand \bibitemStop [0]{}%
\providecommand \bibitemNoStop [0]{.\EOS\space}%
\providecommand \EOS [0]{\spacefactor3000\relax}%
\providecommand \BibitemShut  [1]{\csname bibitem#1\endcsname}%
\let\auto@bib@innerbib\@empty
\bibitem [{\citenamefont {Brown}\ \emph {et~al.}(2018)\citenamefont {Brown},
  \citenamefont {Lacroix}, \citenamefont {Lloyd}, \citenamefont {Boehm},\ and\
  \citenamefont {Chadwick}}]{Brown:2018pwq}%
  \BibitemOpen
  \bibfield  {author} {\bibinfo {author} {\bibfnamefont {A.~M.}\ \bibnamefont
  {Brown}}, \bibinfo {author} {\bibfnamefont {T.}~\bibnamefont {Lacroix}},
  \bibinfo {author} {\bibfnamefont {S.}~\bibnamefont {Lloyd}}, \bibinfo
  {author} {\bibfnamefont {C.}~\bibnamefont {Boehm}}, \ and\ \bibinfo {author}
  {\bibfnamefont {P.}~\bibnamefont {Chadwick}},\ }\href@noop {} {\bibfield
  {journal} {\bibinfo  {journal} {Accepted for publication in PRD}\ } (\bibinfo
  {year} {2018})},\ \Eprint {http://arxiv.org/abs/1806.01866} {arXiv:1806.01866
  [astro-ph.HE]} \BibitemShut {NoStop}%
\bibitem [{\citenamefont {{Abdo}}\ \emph {et~al.}(2010)\citenamefont {{Abdo}}
  \emph {et~al.}}]{2010A&A...524A..75A}%
  \BibitemOpen
  \bibfield  {author} {\bibinfo {author} {\bibfnamefont {A.~A.}\ \bibnamefont
  {{Abdo}}} \emph {et~al.} (\bibinfo {collaboration} {Fermi-LAT}),\ }\href
  {\doibase 10.1051/0004-6361/201014458} {\bibfield  {journal} {\bibinfo
  {journal} {Astron. Astrophys.}\ }\textbf {\bibinfo {volume} {524}},\ \bibinfo
  {pages} {A75} (\bibinfo {year} {2010})},\ \Eprint
  {http://arxiv.org/abs/1003.3588} {arXiv:1003.3588 [astro-ph.GA]} \BibitemShut
  {NoStop}%
\bibitem [{\citenamefont {{Freire}}\ \emph {et~al.}(2011)\citenamefont
  {{Freire}} \emph {et~al.}}]{2011Sci...334.1107F}%
  \BibitemOpen
  \bibfield  {author} {\bibinfo {author} {\bibfnamefont {P.~C.~C.}\
  \bibnamefont {{Freire}}} \emph {et~al.} (\bibinfo {collaboration}
  {Fermi-LAT}),\ }\href {\doibase 10.1126/science.1207141} {\bibfield
  {journal} {\bibinfo  {journal} {Science}\ }\textbf {\bibinfo {volume}
  {334}},\ \bibinfo {pages} {1107} (\bibinfo {year} {2011})},\ \Eprint
  {http://arxiv.org/abs/1111.3754} {arXiv:1111.3754} \BibitemShut {NoStop}%
\bibitem [{\citenamefont {Johnson}\ \emph {et~al.}(2013)\citenamefont {Johnson}
  \emph {et~al.}}]{Johnson:2013uza}%
  \BibitemOpen
  \bibfield  {author} {\bibinfo {author} {\bibfnamefont {T.~J.}\ \bibnamefont
  {Johnson}} \emph {et~al.},\ }\href {\doibase 10.1088/0004-637X/778/2/106}
  {\bibfield  {journal} {\bibinfo  {journal} {Astrophys. J.}\ }\textbf
  {\bibinfo {volume} {778}},\ \bibinfo {pages} {106} (\bibinfo {year}
  {2013})},\ \Eprint {http://arxiv.org/abs/1310.1852} {arXiv:1310.1852
  [astro-ph.HE]} \BibitemShut {NoStop}%
\bibitem [{\citenamefont {Hooper}\ and\ \citenamefont
  {Linden}(2016)}]{Hooper:2016rap}%
  \BibitemOpen
  \bibfield  {author} {\bibinfo {author} {\bibfnamefont {D.}~\bibnamefont
  {Hooper}}\ and\ \bibinfo {author} {\bibfnamefont {T.}~\bibnamefont
  {Linden}},\ }\href {\doibase 10.1088/1475-7516/2016/08/018} {\bibfield
  {journal} {\bibinfo  {journal} {JCAP}\ }\textbf {\bibinfo {volume} {1608}},\
  \bibinfo {pages} {018} (\bibinfo {year} {2016})},\ \Eprint
  {http://arxiv.org/abs/1606.09250} {arXiv:1606.09250 [astro-ph.HE]}
  \BibitemShut {NoStop}%
\bibitem [{\citenamefont {Abdo}\ \emph
  {et~al.}(2009{\natexlab{a}})\citenamefont {Abdo} \emph
  {et~al.}}]{Abdo:2009rjc}%
  \BibitemOpen
  \bibfield  {author} {\bibinfo {author} {\bibfnamefont {A.~A.}\ \bibnamefont
  {Abdo}} \emph {et~al.} (\bibinfo {collaboration} {Fermi-LAT}),\ }\href
  {\doibase 10.1126/science.1177023} {\bibfield  {journal} {\bibinfo  {journal}
  {Science}\ }\textbf {\bibinfo {volume} {325}},\ \bibinfo {pages} {845}
  (\bibinfo {year} {2009}{\natexlab{a}})}\BibitemShut {NoStop}%
\bibitem [{\citenamefont {Xing}\ and\ \citenamefont
  {Wang}(2016)}]{Xing:2016ncp}%
  \BibitemOpen
  \bibfield  {author} {\bibinfo {author} {\bibfnamefont {Y.}~\bibnamefont
  {Xing}}\ and\ \bibinfo {author} {\bibfnamefont {Z.}~\bibnamefont {Wang}},\
  }\href {\doibase 10.3847/0004-637X/831/2/143} {\bibfield  {journal} {\bibinfo
   {journal} {Astrophys. J.}\ }\textbf {\bibinfo {volume} {831}},\ \bibinfo
  {pages} {143} (\bibinfo {year} {2016})},\ \Eprint
  {http://arxiv.org/abs/1604.08710} {arXiv:1604.08710 [astro-ph.HE]}
  \BibitemShut {NoStop}%
\bibitem [{\citenamefont {Abdo}\ \emph
  {et~al.}(2009{\natexlab{b}})\citenamefont {Abdo} \emph
  {et~al.}}]{Abdo:2009jfa}%
  \BibitemOpen
  \bibfield  {author} {\bibinfo {author} {\bibfnamefont {A.~A.}\ \bibnamefont
  {Abdo}} \emph {et~al.} (\bibinfo {collaboration} {Fermi-LAT}),\ }\href
  {\doibase 10.1126/science.1176113} {\bibfield  {journal} {\bibinfo  {journal}
  {Science}\ }\textbf {\bibinfo {volume} {325}},\ \bibinfo {pages} {848}
  (\bibinfo {year} {2009}{\natexlab{b}})}\BibitemShut {NoStop}%
\bibitem [{\citenamefont {Grindlay}\ \emph {et~al.}(2001)\citenamefont
  {Grindlay}, \citenamefont {Heinke}, \citenamefont {Edmonds},\ and\
  \citenamefont {Murray}}]{Grindlay:2001xh}%
  \BibitemOpen
  \bibfield  {author} {\bibinfo {author} {\bibfnamefont {J.~E.}\ \bibnamefont
  {Grindlay}}, \bibinfo {author} {\bibfnamefont {C.}~\bibnamefont {Heinke}},
  \bibinfo {author} {\bibfnamefont {P.~D.}\ \bibnamefont {Edmonds}}, \ and\
  \bibinfo {author} {\bibfnamefont {S.~S.}\ \bibnamefont {Murray}},\ }\href
  {\doibase 10.1126/science.1061135} {\bibfield  {journal} {\bibinfo  {journal}
  {Science}\ }\textbf {\bibinfo {volume} {292}},\ \bibinfo {pages} {2290}
  (\bibinfo {year} {2001})},\ \Eprint {http://arxiv.org/abs/astro-ph/0105528}
  {arXiv:astro-ph/0105528 [astro-ph]} \BibitemShut {NoStop}%
\bibitem [{\citenamefont {Bartels}\ \emph {et~al.}(2018)\citenamefont
  {Bartels}, \citenamefont {Edwards},\ and\ \citenamefont
  {Weniger}}]{Bartels:2018xom}%
  \BibitemOpen
  \bibfield  {author} {\bibinfo {author} {\bibfnamefont {R.~T.}\ \bibnamefont
  {Bartels}}, \bibinfo {author} {\bibfnamefont {T.~D.~P.}\ \bibnamefont
  {Edwards}}, \ and\ \bibinfo {author} {\bibfnamefont {C.}~\bibnamefont
  {Weniger}},\ }\href@noop {} {\  (\bibinfo {year} {2018})},\ \Eprint
  {http://arxiv.org/abs/1805.11097} {arXiv:1805.11097 [astro-ph.HE]}
  \BibitemShut {NoStop}%
\bibitem [{Note1()}]{Note1}%
  \BibitemOpen
  \bibinfo {note} {Similar results were found using a broken power
  law.}\BibitemShut {Stop}%
\bibitem [{\citenamefont {Abdo}\ \emph {et~al.}(2013)\citenamefont {Abdo} \emph
  {et~al.}}]{TheFermi-LAT:2013ssa}%
  \BibitemOpen
  \bibfield  {author} {\bibinfo {author} {\bibfnamefont {A.~A.}\ \bibnamefont
  {Abdo}} \emph {et~al.} (\bibinfo {collaboration} {Fermi-LAT}),\ }\href
  {\doibase 10.1088/0067-0049/208/2/17} {\bibfield  {journal} {\bibinfo
  {journal} {Astrophys. J. Suppl.}\ }\textbf {\bibinfo {volume} {208}},\
  \bibinfo {pages} {17} (\bibinfo {year} {2013})},\ \Eprint
  {http://arxiv.org/abs/1305.4385} {arXiv:1305.4385 [astro-ph.HE]} \BibitemShut
  {NoStop}%
\bibitem [{\citenamefont {Cirelli}\ \emph {et~al.}(2011)\citenamefont
  {Cirelli}, \citenamefont {Corcella}, \citenamefont {Hektor}, \citenamefont
  {Hutsi}, \citenamefont {Kadastik}, \citenamefont {Panci}, \citenamefont
  {Raidal}, \citenamefont {Sala},\ and\ \citenamefont
  {Strumia}}]{Cirelli:2010xx}%
  \BibitemOpen
  \bibfield  {author} {\bibinfo {author} {\bibfnamefont {M.}~\bibnamefont
  {Cirelli}}, \bibinfo {author} {\bibfnamefont {G.}~\bibnamefont {Corcella}},
  \bibinfo {author} {\bibfnamefont {A.}~\bibnamefont {Hektor}}, \bibinfo
  {author} {\bibfnamefont {G.}~\bibnamefont {Hutsi}}, \bibinfo {author}
  {\bibfnamefont {M.}~\bibnamefont {Kadastik}}, \bibinfo {author}
  {\bibfnamefont {P.}~\bibnamefont {Panci}}, \bibinfo {author} {\bibfnamefont
  {M.}~\bibnamefont {Raidal}}, \bibinfo {author} {\bibfnamefont
  {F.}~\bibnamefont {Sala}}, \ and\ \bibinfo {author} {\bibfnamefont
  {A.}~\bibnamefont {Strumia}},\ }\href {\doibase
  10.1088/1475-7516/2012/10/E01, 10.1088/1475-7516/2011/03/051} {\bibfield
  {journal} {\bibinfo  {journal} {JCAP}\ }\textbf {\bibinfo {volume} {1103}},\
  \bibinfo {pages} {051} (\bibinfo {year} {2011})},\ \bibinfo {note} {[Erratum:
  JCAP1210,E01(2012)]},\ \Eprint {http://arxiv.org/abs/1012.4515}
  {arXiv:1012.4515 [hep-ph]} \BibitemShut {NoStop}%
\bibitem [{\citenamefont {{Wood}}\ \emph {et~al.}(2017)\citenamefont {{Wood}},
  \citenamefont {{Caputo}}, \citenamefont {{Charles}}, \citenamefont {{Di
  Mauro}}, \citenamefont {{Magill}},\ and\ \citenamefont {{Jeremy Perkins for
  the Fermi-LAT Collaboration}}}]{2017arXiv170709551W}%
  \BibitemOpen
  \bibfield  {author} {\bibinfo {author} {\bibfnamefont {M.}~\bibnamefont
  {{Wood}}}, \bibinfo {author} {\bibfnamefont {R.}~\bibnamefont {{Caputo}}},
  \bibinfo {author} {\bibfnamefont {E.}~\bibnamefont {{Charles}}}, \bibinfo
  {author} {\bibfnamefont {M.}~\bibnamefont {{Di Mauro}}}, \bibinfo {author}
  {\bibfnamefont {J.}~\bibnamefont {{Magill}}}, \ and\ \bibinfo {author}
  {\bibnamefont {{Jeremy Perkins for the Fermi-LAT Collaboration}}},\
  }\href@noop {} {\bibfield  {journal} {\bibinfo  {journal} {ArXiv e-prints}\ }
  (\bibinfo {year} {2017})},\ \Eprint {http://arxiv.org/abs/1707.09551}
  {arXiv:1707.09551 [astro-ph.IM]} \BibitemShut {NoStop}%
\bibitem [{\citenamefont {Freire}\ \emph {et~al.}(2017)\citenamefont {Freire}
  \emph {et~al.}}]{Freire:2017mgu}%
  \BibitemOpen
  \bibfield  {author} {\bibinfo {author} {\bibfnamefont {P.~C.~C.}\
  \bibnamefont {Freire}} \emph {et~al.},\ }\href {\doibase
  10.1093/mnras/stx1533} {\bibfield  {journal} {\bibinfo  {journal} {Mon. Not.
  Roy. Astron. Soc.}\ }\textbf {\bibinfo {volume} {471}},\ \bibinfo {pages}
  {857} (\bibinfo {year} {2017})},\ \Eprint {http://arxiv.org/abs/1706.04908}
  {arXiv:1706.04908 [astro-ph.HE]} \BibitemShut {NoStop}%
\bibitem [{Note2()}]{Note2}%
  \BibitemOpen
  \bibinfo {note} {The conversion from p-value ($\alpha $) to $\sigma $ is
  performed using the inverse-cumulative-distribution function for a
  standard-normal distribution: $\protect \mathrm {CDF}^{-1}_\protect \mathcal
  {N}\left (1-\alpha \right )$.}\BibitemShut {Stop}%
\end{thebibliography}%

\end{document}